\documentclass[aps,twocolumn,showpacs,floats,superscriptaddress]{revtex4}
\usepackage{graphicx}
\usepackage{dcolumn}
\usepackage{bm}
\begin{document}

\author{N.V. Prokof'ev}
\affiliation{Department of Physics, University of Massachusetts,
Amherst, MA 01003, USA}
\affiliation{Theoretische Physik, ETH Z\"{u}rich, CH-8093 Z\"{u}rich, Switzerland}
\affiliation{Russian Research Center
``Kurchatov Institute'', 123182 Moscow, Russia}

\author{B.V. Svistunov}
\affiliation{Department of Physics, University of Massachusetts,
Amherst, MA 01003, USA} \affiliation{Russian Research Center
``Kurchatov Institute'', 123182 Moscow, Russia}

\title{Bold diagrammatic Monte Carlo:\\
A generic technique for polaron (and many-body?) problems}

\begin{abstract}
We develop a Monte Carlo scheme for sampling series of Feynman diagrams
for the proper self-energy which are self-consistently expressed in terms
of renormalized particle propagators. This approach is used to
solve the problem of a single spin-down fermion resonantly interacting
with the Fermi gas of spin-up particles. Though the original series
based on bare propagators are sign-alternating and divergent one can
still determine the answer behind them by using two strategies
(separately or together):
(i) using proper series re-summation techniques, and
(ii) introducing renormalized propagators which are defined in terms of the
simulated proper self-energy, i.e. making the entire scheme self-consistent.
Our solution is important for understanding
the phase diagram and properties of the BCS-BEC crossover in the strongly
imbalanced regime.
On the technical side, we develop a generic sign-problem tolerant
method for exact numerical solution of polaron-type models, and, possibly,
of the interacting many-body Hamiltonians.
\end{abstract}

  \maketitle

\section{Introduction}
\label{sec:I}

Modern science has radically changed our view of the physical
vacuum. Instead of na\"ive ``empty space'' we have to deal with a
complex groundstate of an interacting system, and, strictly
speaking, there is no fundamental difference between the outer
space, helium, or any other condensed matter system. With this
point of view comes understanding that the notion of a ``bare''
particle is somewhat abstract since its coupling to vacuum
fluctuations, or environment, may strongly (sometimes radically)
change particle properties at energies addressed by the
experimental probes. The polaron problem  \cite{Landau} is by now
canonical across all of physics with the same questions about
particle energy, effective mass, residue, etc., being asked for
different types of particles, environments and coupling between
them \cite{Appel}. In a broader context, particles are not
necessarily external objects unrelated to a given vacuum, but
quasiparticle excitations of the very same ground state. Thus, the
solution of the polaron problem paves the way to the effective low
energy theory of a given system, and, to large extent, determines
basic properties of all condensed matter systems at low
temperature.

At the moment, there is no generic analytic or numeric technique
to study quasiparticle properties for arbitrary strongly
interacting system. Analytic solutions are typically (with few
exceptions, see, e.g., \cite{Zotos}) based on perturbative
corrections to certain limiting cases
\cite{Landau,Frohlich,Schultz,Lee,Appel} or variational treatment
\cite{Feynman}. Several numeric schemes were suggested in the
past, but all of them have limitations either in the system size,
system dimension, interaction or environment type. Exact
diagonalization and variational methods in the low-energy subspace
\cite{Bonca,Fehske} are mostly restricted to one-dimensional
models with short-range interactions. The continuous time path
integral formulation \cite{Kornilovitch} works for the lattice
model with linear coupling between the particle and bosonic
environment, but it can not be generalized to fermionic
environment, sign-alternating coupling (as in the $t$-$J$ model
\cite{tJ}), nor is it suitable for continuous-space models.

In this article (which follows a short communication
\cite{ourPRB}), we develop a Monte Carlo technique which simulates
series of Feynman diagrams for the proper self-energy. The
diagrammatic Monte Carlo (Diag-MC) technique was used previously
to study electron-phonon polarons \cite{polaron98,polaron2000}.
The essence of Diag-MC is in interpreting the sum of all Feynman
diagrams as an ensemble averaging procedure over the corresponding
configuration space. It was considered essential for the method to
work that the series of diagrams for the Green's function be
convergent and sign-positive. Though the configuration space of
diagrams for polarons in Fermi systems is more complex, similar
methods of generating and sampling the corresponding configuration
space can be used. The crucial difference is that in the Fermi
system we have to deal with the sign-alternating and divergent (at
least for strong coupling) series. Under these conditions a direct
summation of all relevant Feynman diagrams for the Green's
functions is not possible, and one has to develop additional tools
for (i) reducing the number of diagrams by calculating
self-energies rather than Green's functions, (ii) employing the
``bold-line'' trick in the form of the Dyson equation which allows
self-consistent summation of infinite geometric series and further
reduces the number of self-energy diagrams, and, if necessary,
(iii) extrapolating Diag-MC results to the infinite diagram order
for a divergent series. At the moment we do not see any obvious
limitations of the new method since even divergent sign
alternating series can be dealt with to obtain reliable results.
We believe that our findings are important in a much broader
context since the Diag-MC approach to the many-body problem has
essentially the same structure.

As a practical application of the method we consider a particle
coupled to the ideal Fermi sea via a short-range potential with
divergent $s$-wave scattering length. This problem was recognized
as the key one for understanding the phase diagram of the
population imbalanced resonant Fermi gas \cite{Bulgac,Lobo}. In
particular, to construct the energy functional describing dilute
solutions of minority (spin down) fermions resonantly coupled to
the majority (spin up) fermions one has to know precisely the
quasiparticle parameters of spin-down fermions since they
determine the coefficients in the energy expansion in the
spin-down concentration $x_{\downarrow}$: the linear term is
controlled by the polaron energy, and the $x_{\downarrow}^{5/3}$
term is determined by the polaron mass \cite{Lobo}.

The general Hamiltonian we deal with in this article can be
written as
\begin{equation}
H\, =\, H_F\,  -\,  \Delta_R/2m\,  + \int d \bm{r} V(\bm{r} -
\bm{R})\: n(\bm{r}) \; , \label{H}
\end{equation}
where $H_F$ is the Hamiltonian of the ideal spin-up Fermi gas with
density $n$ and Fermi momentum $k_F$, $\bm{R}$ is the particle
coordinate, and $ V(\bm{r} - \bm{R})$ is the interaction potential
of finite range $r_0$ between the particle and the spin-up Fermi
gas. In what follows we refer to (\ref{H}) as the Fermi-polaron
problem. The specifics of  the BCS-BEC crossover physics in the
strongly imbalanced regime is two-fold: one is that the particle
and the Fermi gas have the same bare mass $m$ (in what follows we
will use units such that $m=1/2$ and $k_F=1$), and the other is
that one has to take explicitly the so-called zero-range resonant
limit when $r_0 \to 0$, but the $s$-scattering length $a$ remains
finite, i.e. $ k_Fa $ remains fixed for $k_F r_0 \to 0$. In this
limit, the nature of the interaction potential is irrelevant, and
the same results will be obtained e.g. for the neutron matter and
Cesium atoms. We note, however, that the method we develop for the
numeric solution of the resonant Fermi-polaron problem is
absolutely general and can be used for an arbitrary model
described by Eq.~(\ref{H}).

It turns out that the structure of the phase diagram is very
sensitive to polaron parameters. If the state with a dilute gas of
spin-down fermions is stable at all values of $ k_Fa$ then the
solution of the single-particle problem would define the phase
diagram in the vicinity of the multicritical point discussed
recently by Sachdev and Yang \cite{Sachdev}, where four different
phases meet. At this point the spin-down fermion forms a bound
state with a spin-up fermion thus becoming a spin-zero composite
boson (``molecule"); i.e. quasi-particles radically change their
statistics. The multicritical point, however, may be
thermodynamically unstable if the effective scattering length
between the composite bosons and spin-up electrons is large
enough, and the analysis of Refs.~\cite{Carlson,Giorgini} based on
the fixed-node Monte Carlo simulations finds evidence in favor of
this scenario. Phase separation was also found in calculations
based on mean-field and narrow-resonance approaches (both at
finite and zero temperature) see e.g.
Refs.~\cite{Leo,Pao,Iskin,Hu,Chien,Princeton}, though with strong
quantitative deviations from results based on the fixed-node Monte
Carlo simulations \cite{Giorgini}. On the experimental side, MIT
experiments \cite{Shin} are in good agreement with the predictions
made in Ref.~\cite{Giorgini}, while Rice experiments
\cite{Partridge} are not. The origin of discrepancy between the
two experiments is not understood. Our results for polaron
energies are in excellent agreement with Ref.~\cite{Giorgini}.

It is worth noting that the model (\ref{H})
(in general, the particle mass is different from that of the Fermi gas)
is also known as the Anderson orthogonality problem with recoil
\cite{Anderson,Kondo}. It can be also considered as a specific example
of a particle coupled to the Ohmic dissipative bath (see monograph
\cite{Weiss} for numerous other examples and connections to realistic
systems).

The paper is organized as follows. In Sec.~\ref{sec:diagrams} we
discuss the configuration space of Feynman diagrams for
self-energy in momentum--imaginary-time representation (both in
the particle and molecule channels), and explain how polaron
parameters can be obtained in this representation. In
Sec.~\ref{sec:scheme} we describe a Monte Carlo algorithm for
generating and sampling the corresponding diagrammatic space. A
small technical Section \ref{sec:gamma} deals with numerically
evaluating the effective $T$-matrix  by bold diagrammatic Monte
Carlo. We present and discuss results in Sec.~\ref{sec:results}.
In particular, we show that one can use re-summation techniques
for divergent series of diagrams based on bare propagators to
determine the final answer. In Sec.~\ref{sec:bold} we further
advance the algorithm by employing bold-line approach in which the
entire scheme is self-consistently based on renormalized
(``bold-line'') propagators.  We present our conclusions and
perspectives for future work in Sec.~\ref{sec:colclusions}.
\section{Configuration space of self-energy diagrams}
\label{sec:diagrams}

As mentioned above, when coupling between spin-down and spin-up
fermions is strong enough they from a composite boson, or molecule
state. In what follows, we will be using the term ``polaron" in a
narrow sense, i.e. only for the unbound fermionic spin-down
excitation. For the composite boson we will be using the term
``molecule". Since our goal is to calculate particle properties
for arbitrary coupling strength we have to consider one- and
two-particle channels on equal footing. In the rest of this
section we render standard diagrammatic rules for irreducible
self-energy diagrams in both channels, with an emphasis on
specifics of working in the imaginary-time representation.

\subsection{Polaron channel}
We start from the definition of the single-particle Green's
function (see e.g. \cite{book})
\begin{equation}
G(\tau, {\bf r}) = - \langle T_{\tau} \psi (\tau, {\bf r}) \bar{\psi}(0) \rangle
\; , \label{G1}
\end{equation}
and its frequency-momentum representation
\begin{equation}
G(\xi, {\bf p}) = \int {\rm e}^{i(\xi\tau - {\bf p}\cdot
{\bf r})}\, G(\tau , {\bf r})\,  d{\bf r} d\tau  \; . \label{G3}
\end{equation}
Here $\psi (\tau, {\bf r})$ is the fermion annihilation operator
at the space-time point $(\tau, {\bf r})$.
For the ideal spin-up Fermi gas at $T=0$ we have
\begin{equation}
G_{\uparrow}(\xi, {\bf p}) = {1\over i\xi-p^2/2m+\epsilon_F} \; .
\label{G_ideal}
\end{equation}

\begin{figure}[tbp]
\centerline{\includegraphics [bb=70 360 510 430,
width=\columnwidth]{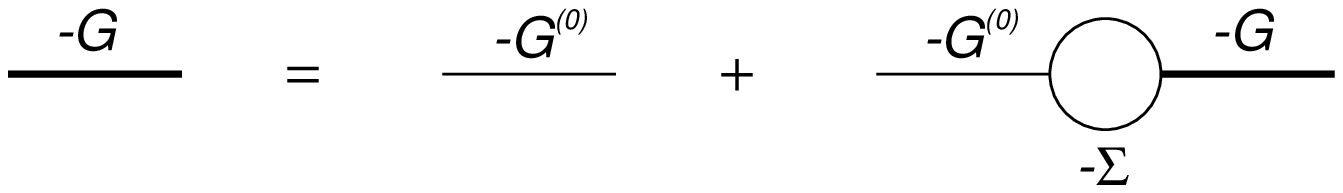}} \caption{Dyson equation for the
single-particle Green's function.} \label{Dyson}
\end{figure}

The vacuum Green's function for the spin-down polaron is
\begin{equation}
G^{(0)}_{\downarrow}(\tau, {\bf p},\mu)\; =\;-\theta(\tau)\, {\rm
e}^{-(p^2/2m-\mu)\tau} \; , \label{G_vac}
\end{equation}
where $\theta$ is the step function, and $\mu$ is a free parameter.
From Dyson's equation for the polaron, see Fig.~\ref{Dyson}, one finds
\begin{equation}
G_{\downarrow}(\xi, {\bf p}, \mu)\; =\; {1\over
i\xi-p^2/2m+\mu-\Sigma(\xi,{\bf p}, \mu)} \; , \label{GG}
\end{equation}
where self-energy $\Sigma$ is given by the sum of all irreducible
diagrams (i.e. diagrams which can not be made disconnected but cutting
through the $G^{(0)}_{\downarrow}$ line) taken with the negative sign.
Taking into account that in the $\tau$-representation both $G{\downarrow}$ and $\Sigma$
depend on $\mu$ only through exponential factors $\exp{(\mu \tau)}$,
one obtains $G_{\downarrow}(\xi, {\bf p}, \mu)\equiv
G_{\downarrow}(\xi-i\mu, {\bf p})$ and $\Sigma(\xi, {\bf p},
\mu)\equiv \Sigma(\xi-i\mu, {\bf p})$.

If polaron is a well-defined quasi-particle, then its energy $E({\bf p})$
and residue $Z({\bf p})$ can be extracted from the asymptotic
decay
\begin{equation}
G_{\downarrow}(\tau, {\bf p}, \mu) \to - Z{\rm
e}^{-(E-\mu)\tau}\; , ~~~~~~~\tau  \to \infty \; \; .
\label{asym}
\end{equation}
This asymptotic behavior immediately implies that the function
$G_{\downarrow}(\xi-i\mu, {\bf p})$ has a pole singularity
\begin{equation}
G_{\downarrow}(\xi-i\mu, {\bf p}) = {Z({\bf p})\over i\xi +\mu
-E({\bf p})}  + {\rm regular~part} \; . \label{pole}
\end{equation}
Now setting  $\mu=E({\bf p})$ in (\ref{pole}) and comparing the
result to (\ref{GG}), we conclude that
\begin{equation}
i \xi / Z = i\xi -p^2/2m + E - \Sigma(0,{\bf p}, E) + i\xi
A({\bf p},E)\; , \label{rel1}
\end{equation}
where (we take into account that
$\partial \Sigma /\partial \xi = i \partial \Sigma /\partial \mu $)
\begin{equation}
A({\bf p},E) = - \left. {\partial  \Sigma (0,{\bf p}, \mu)
\over
\partial \mu}\, \right\vert_{\mu=E}\; . \label{A}
\end{equation}
This yields two important relations (see also \cite{book}):
\begin{equation}
E = p^2/2m + \Sigma (0,{\bf p}, E) \; , \label{E}
\end{equation}
and
\begin{equation}
Z = {1\over 1+ A({\bf p},E)}\; . \label{Z}
\end{equation}
Equation~(\ref{E}) allows one to solve for $E$
provided $\Sigma(\tau, {\bf p}, \mu)$ is known.
All we have to do is to calculate the the zero-frequency value
of $\Sigma$ for $\mu=E$
\begin{equation}
E = p^2/2m + \int_0^{\infty} \Sigma (\tau,{\bf p}, \mu)\,
{\rm e}^{(E-\mu)\tau}\, d\tau  \; . \label{EE}
\end{equation}
After $E$ is found, the residue is obtained from Eq.~(\ref{Z})
using
\begin{equation}
A({\bf p},E)= -\int_0^{\infty}\tau  \Sigma (\tau,{\bf p}, \mu)\, {\rm
e}^{(E-\mu)\tau}\, d\tau  \; . \label{AA}
\end{equation}
Note also that the dependence on $\mu$ drops out from both
(\ref{EE}) and (\ref{AA}).

A comment is in order here. Strictly speaking,  polaron and molecule
poles exist only for ${\bf p}=0$ because the fermionic bath they couple to
is gapless. However, the spectrum $E({\bf p})$ is well-defined in the limit
$p\to 0$, as the decay width vanishes faster than $[E({\bf
p})-E(0)]$. To have stable quasi-particles, one can use a trick of
introducing a gap $\Delta$ in the environment spectrum, e.g. by redefining the
dispersion relation for spin-up fermions:
$\varepsilon_{\bf k}\to \max(\varepsilon_{\bf k},\Delta)$.
In the $p\to 0$ limit, the systematic error vanishes
faster than $[E({\bf p})-E(0)]$, provided $\Delta \sim [E({\bf
p})-E(0)]$. It should be also possible to work with $\Delta $'s
essentially larger than $[E({\bf p})-E(0)]$ and extrapolate to
$\Delta \to 0$. In particular, such an extrapolation is possible
(and is implicitly implied) at the analytical level in the relation
for the effective mass, which we consider below.

One way to determine the effective mass is
to calculate the quasi-particle energy as a function of momentum
for a set of small but finite values of $p$ and fit it with the parabola.
It is, however, possible to skip this procedure and to write a direct
estimator for the effective mass in terms of the calculated self-energy.
Acting with the operator $\nabla_{\bf P}^2$
on both sides of Eq.~(\ref{E}) and taking the limit $p\to 0$
we get
\begin{equation}
{1+A_0\over m_*} = {1\over m} + B_0 \; , \label{m_*}
\end{equation}
\begin{equation}
B_0 =  {1 \over 3} \int_0^{\infty} d\tau\, {\rm
e}^{(E_0-\mu)\tau} \left. [ \nabla_{\bf P}^2\, \Sigma (\tau,{\bf
p}, \mu)] \right\vert_{p=0} \; , \label{B_0}
\end{equation}
where $A_0\equiv A({\bf p}=0)$ and $E_0\equiv E({\bf p}=0)$.

\begin{figure}[tbp]
\centerline{\includegraphics [bb=10 360 390 450,
width=\columnwidth]{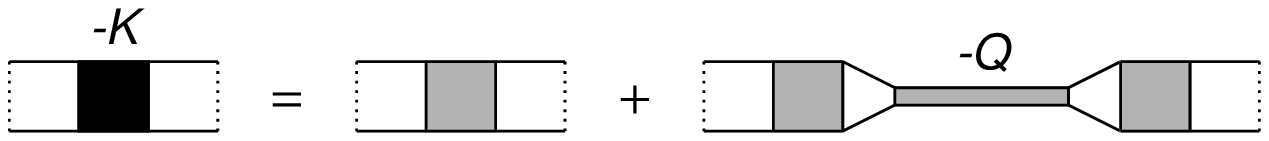}} \centerline{\includegraphics [bb=0 360 450
430, width=\columnwidth]{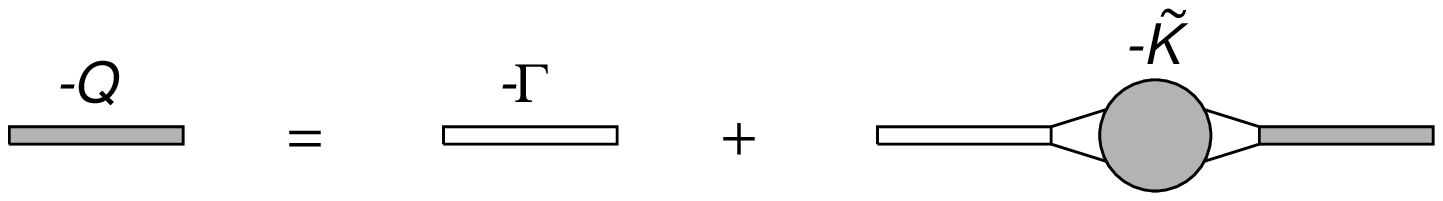}} \caption{Defining functions $Q$ and
$\tilde{K}$ in the two-particle channel. } \label{K}
\end{figure}

\subsection{Molecule channel}
In this case we start with the two-particle propagator
\begin{equation}
K(\tau, {\bf p}) = -\langle \, {\rm T}_\tau \Phi_{\bf
p}(\tau)\, \Phi_{\bf p}^{\dagger}(0)\, \rangle \; , \label{001}
\end{equation}
where
\begin{equation}
\Phi_{\bf p} = \int {d{\bf q}\over 2\pi}\, \varphi_{\bf q} \,
\psi_{\uparrow }({\bf p}-{\bf q}) \psi_{\downarrow} ({\bf q}) \; ,
\label{002}
\end{equation}
and $\varphi_{\bf q}$ is the pair wavefunction (in momentum
representation) that localizes the relative distance between two particles.
If there is a bound state (molecule), then
\begin{equation}
K(\tau, {\bf p}, \mu) \to - Z_{\rm mol}\, {\rm e}^{-(E_{\rm
mol}-\mu)\tau}\; , ~~\tau  \to \infty \; , \label{asym_K}
\end{equation}
and the pair propagator in the frequency representation
has a pole:
\begin{equation}
K(\xi-i\mu, {\bf p}) = {Z_{\rm mol}({\bf p})\over i\xi +\mu
-E_{\rm mol}({\bf p})}  + {\rm regular~part} \; .
\label{pole_K}
\end{equation}

Now we introduce yet another function, that features the same
molecular pole, but has a simpler diagrammatic structure. The
specifics of the resonant zero-range limit is that the sum of all
ladder diagrams for the interaction potential $V(r)$ has to be
considered as a separate diagrammatic element. We denote this sum
by  $\Gamma (\tau ,p)$ and consider it as a ``bare pair
propagator''. Of course, the same approach can be taken in a
general case to replace the bare interaction potential with the
scattering $T$-matrix, but in the zero-range limit we really do
not have any other alternative for the
ultra-violet-divergence-free formulation. The sum of ladder
diagrams takes the ultra-violet physics into account exactly and
allows to express $\Gamma(\tau ,p)$ in terms of the $s$-scattering
length $a$. The ladder structure of diagrams absorbed in
$\Gamma(\tau ,p)$ also explains why we treat it as a ``pair
propagator'' (we will depict it with a double-line, see Fig~\ref{K}). The
exact expression for $\Gamma$ is readily obtained from the
geometrical series (in frequency domain)
\begin{equation}
-\Gamma\, =\, -U\,  +\, (-U)^2\, \Pi\,  +\,  \dots \; ,
\label{ladder}
\end{equation}
where the polarization operator is defined by
\begin{equation}
\Pi(\eta, p) = \int_{q > k_F} { d{\bf q} /(2\pi)^3 \over
q^2/2m+({\bf p}-{\bf q})^2/2m -\eta  }\; , \label{aa}
\end{equation}
and $ \eta = \omega + \varepsilon_F+\mu$. Using suitable
ultra-violet regularization, one can cast the same expression in
the universal form which depends only on the $s$-scattering
length:
\begin{equation}
\Gamma^{-1}(\eta,p) ={m\over 4\pi a} - {m\over 8\pi
}\sqrt{p^2-4m\eta}-\bar{\Pi }(\eta, p)
 \; , \label{Gamma}
\end{equation}
\begin{equation}
\bar{\Pi}(\eta, p) = \int_{q\leq k_F} { d{\bf q} /(2\pi)^3 \over
q^2/2m+({\bf p}-{\bf q})^2/2m -\eta  }\; . \label{a}
\end{equation}
For finite density of spin-up fermions converting
Eq.~(\ref{Gamma}) to the imaginary time domain has to be done
numerically. One possibility is to use the inverse Laplace
transform. We employed the bold diagrammatic Monte Carlo
technology \cite{BMC} to achieve this goal and further details are
provided in Sec.~\ref{sec:gamma}. The two-dimensional function
$\Gamma(\tau,p)$ is tabulated prior to the polaron simulation.

In Fig.~\ref{K} we define function $Q$ that can be viewed as a
renormalized pair propagator related to $\Gamma$ by the Dyson
equation. In the upper panel, we show the diagrammatic structure
for $K$, which includes dotted lines representing external
functions $\varphi_{\bf q}$, grey squares representing sums of all
$\Gamma$-irreducible diagrams, and the renormalized pair
propagator $Q$. By $\Gamma$-irreducible diagrams we understand
diagrams which can not be made disconnected by cutting them
through a single $\Gamma$-line. All $\Gamma$-reducible diagrams
are absorbed in the $Q$ function which is shown in the lower panel
in Fig.~\ref{K}. The grey circle has nearly the same structure as
the grey square (the zeroth order term is present in the crossed
square, but not in the crossed circle): since $\Gamma$ is defined
as the sum of ladder diagrams, all terms which include ladder-type
structures based on free one-particle propagators have to be
excluded from $Q$ and $K$. With the replacements $G_{\downarrow}
\to Q$, $G^{(0)}_{\downarrow} \to \Gamma$, and $\Sigma \to
\tilde{K}$ we find an exact analogy between the one- and
two-particle propagators.

The analogy can be carried out further by noting that
the structure of diagrams in Fig.~\ref{K} implies that $Q$ has
the same pole as $K$, while the rest of the functions simply change the
value of the quasi-particle residue. Thus, if molecule is a
well-defined excitation we expect that
\begin{equation}
Q(\xi-i\mu, {\bf p}) = {\tilde{Z}_{\rm mol}({\bf p})\over i\xi
+\mu -E_{\rm mol}({\bf p})}  + {\rm regular~part} \; .
\label{pole_Q}
\end{equation}
This explains why introducing the function $Q$ is convenient: now
Eqs.~(\ref{Z})-(\ref{B_0}) are immediately generalized to the
molecule case (up to replacements mentioned above).

\section{Worm algorithm for Feynman diagrams}
\label{sec:scheme}

In this Section we describe how the configuration space of Feynman diagrams
for $\Sigma$ and $\tilde{K}$ is parameterized and updated using Diag-MC
rules. Our algorithm is designed to treat polaron and molecule channels on
equal footing. We achieve this goal by introducing auxiliary diagrams which
contain two ``loose'' ends called ``worms''---this was proven to be an
efficient strategy for reducing the MC autocorrelation time when simulations
are performed in the configuration space with complex topology
\cite{worms,worms2}.

\subsection{Cyclic diagrams}
\begin{figure}[tbp]
\centerline{\includegraphics [bb=80 360 440 450,
width=\columnwidth]{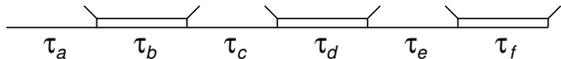}} \caption{The backbone of the cyclic
diagram.} \label{pre_diag}
\end{figure}

\begin{figure}[tbp]
\centerline{\includegraphics [bb=80 360 440 450,
width=\columnwidth]{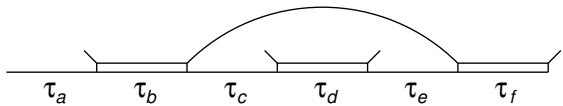}} \caption{Forward connection. The arc
represents $-G_{\uparrow}(\tau=\tau_c+\tau_d+\tau_e)$.}
\label{forward}
\end{figure}

\begin{figure}[tbp]
\centerline{\includegraphics [bb=80 360 440 450,
width=\columnwidth]{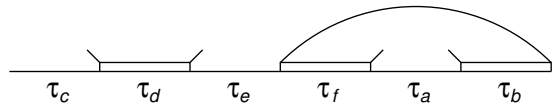}} \caption{Backward connection. The
pair of $\Gamma$-ends being connected is precisely the same as in
Fig.~\ref{forward}, but the direction is opposite, and the arc
represents  $-G_{\uparrow}(\tau=-\tau_f -\tau_a-\tau_b)$.}
\label{backward}
\end{figure}

We start with the introduction of cyclic diagrams. Though we work
in the imaginary time representation at $T=0$ when $\tau \in
[0,\infty)$, it is still convenient not to specify the time origin
and to consider diagrams on the imaginary time circle.  The
backbone of each cyclic diagram is a pre-diagram illustrated in
Fig.~\ref{pre_diag}. It consists of a cyclic chain of the
structure $~G^{(0)}_{\downarrow}(\tau_a)\, \Gamma(\tau_b) \,
G^{(0)}_{\downarrow}(\tau_c)\, \Gamma(\tau_d) \,
G^{(0)}_{\downarrow}(\tau_e) \, \Gamma(\tau_f)\, \ldots \,  $ (all
the times are positive). We do not explicitly show ``directions"
of the propagators, since these are unambiguously fixed by the
global direction of all the backbone lines, which we
select---without loss of generality---to be from right to left.
With this convention, the left (free) spin-up end of any
$\Gamma$-line is outgoing, while the right end is incoming. A
physical diagram is obtained by pairwise replacing free spin-up
ends with propagators $G_{\uparrow}$. There are two ways to
connect incoming and outgoing lines: (i) forward, i.e., in the
direction of the backbone propagators, and (ii) backward (opposite
to forward). Forward (backward) connections result in propagators
$G_{\uparrow}$ with positive (negative) times, see
Figs.~\ref{forward} and \ref{backward} for illustrations. They
represent particle (hole) excitations in the fermionic
environment. It is important to emphasize that in cyclic diagrams
the only time-variables are the positive time-lengths of
$G^{(0)}_{\downarrow}$'s and $\Gamma$'s. There is no absolute
time, and, correspondingly, all moments in time are equivalent.


\subsection{Worms}
\begin{figure}[tbp]
\centerline{\includegraphics [bb=60 360 440 450,
width=\columnwidth]{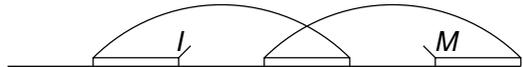} } \caption{A diagram with two worms,
${\cal I}$ and ${\cal M}$.} \label{Worms}
\end{figure}

\begin{figure}[tbp]
\centerline{\includegraphics [bb=70 360 420 450,
width=\columnwidth]{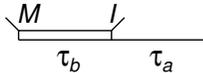} } \caption{``Normalization" diagram.
It is the simplest diagram with the worm; its weight is a product
of $G^{(0)}_\downarrow(\tau_a)$ and $\Gamma(\tau_b)$.}
\label{zeroth}
\end{figure}

To have a MC scheme which simulates diagrams in one- and
two-particle channels on equal footing we extend the space of
physical diagrams by allowing diagrams with two special
end-points, or ``worms''. They will be denoted ${\cal I}$ and
${\cal M}$ and stand for the unconnected incoming (outgoing)
spin-up ends, see Fig.~\ref{Worms} for an illustration.
Correspondingly, the entire updating scheme is based on
manipulations with ${\cal I}$ and ${\cal M}$. As it will become
clear soon, of special importance for normalization of MC results
is the first-order diagram with the worm, see Fig.~\ref{zeroth}.
Its weight consists of just two factors, $G^{(0)}_{\downarrow}(\tau_a)$
and $\Gamma(\tau_b)$.

\subsection{Parametrization of diagrams}
Apart from the diagram order and its topology, we select time
intervals of $\Gamma$'s and $G_\downarrow^{(0)}$'s, and momenta of
the spin-up propagators as independent variables. The momenta of
$\Gamma$'s and $G_\downarrow^{(0)}$'s are then unambiguously
defined by the momentum conservation law, while the time interval
of a spin-up propagator is obtained by summing up the time
intervals of $\Gamma$'s and $G_\downarrow^{(0)}$'s covered by this
propagator. Technically, we find it convenient to work explicitly
in the particle-hole representation for the spin-up propagators
when backward spin-up propagator is understood as a forward hole
propagator with the opposite momentum. This is achieved by
introducing a non-negative function,
\begin{equation}
\tilde{G}(\tau, {\bf p}) = \left\{
\begin{array}{l}
-G_\uparrow(\tau, {\bf p})\; , ~~ p\, \geq \, p_F \; , \\
G_\uparrow(-\tau, -{\bf p})\; , ~~ p\, < \, p_F\; ,\end{array}
\right.  \label{tilde_G}
\end{equation}
which is assigned to all spin-up lines (the global fermionic sign of
the diagram is defined separately, by standard diagrammatic
rules). All momenta assigned to the spin-up lines are
understood as momenta of the corresponding
$\tilde{G}$-propagators; i.e. they are either momenta of
particles (for forward propagators they are non-zero only for
$p\geq p_F$), or momenta of holes (for backward propagators they
are non-zero only for $p<p_F$). An explicit formula for
$\tilde{G}$ (subject to the above conditions) is
\begin{equation}
\tilde{G}(\tau, {\bf p}) =\theta(\tau)\, {\rm
e}^{-|p^2/2m-\varepsilon_F|\tau} \; . \label{tilde_G_expl}
\end{equation}

To simplify the description of updates below we will generically
refer to $G^{(0)}_{\downarrow}$- and $\Gamma$-propagators as
backbone lines (BBLs) and denote them as $D$. The diagram order
$N$ is given by the total number of spin-down propagators. We also
use special counters to characterize the topology of the diagram.
For each BBL we define a {\it cover number}, $N_c$, equal to the
total number of $\tilde{G}$-lines covering a given BBL. A backbone
line with $N_c=0$ is called uncovered. Finally, physical
diagrams---the ones without worms---are divided into polaron (0)
and molecule (1) sectors; the diagram sector is defined by the
difference between the number of particle and hole spin-up
propagators covering any of the $G^{(0)}_{\downarrow}$-lines (the
same result is obtained by analyzing propagators covering
$\Gamma$-lines after adding unity for the spin-up particle
participating in the ladder diagrams).

\subsection{Updates}

The cyclic structure of diagrams in combination with the
possibility of considering non-physical diagrams allows one to
construct a very simple ergodic set of updates. The minimal set
consists of two complementary pairs, {\it Insert/Delete} and {\it
Open/Close}, and one self-complementary update {\it Reconnect}.
The description that follows aims at the most transparent and
straightforward realization of updates. Standard performance
enhancement tricks and optimization protocols are not mentioned.
In particular, we base our considerations on the updating scheme
which may propose a change leading to a forbidden configuration.
Such proposals are rejected as if they result in zero acceptance
ratio.

 {\bf Insert:} This update applies only to physical---no
worms---diagrams and is rejected otherwise. First, consider the
polaron sector. Select at random one of the $G^{(0)}_\downarrow$
propagators; if it is covered, the update is automatically
rejected. If the selected propagator is uncovered, insert a pair
of new propagators, $\Gamma(\tau_1, {\bf p})$ and
$G^{(0)}_\downarrow(\tau_2, {\bf p})$, right after the selected
one. The new $\Gamma(\tau_1, {\bf p})$-propagator is supposed to
contain ${\cal I}$ and ${\cal M}$ at its ends. Worms radically
simplify this diagram-order increasing update since due to
conservation laws the momenta of new BBLs are equal to the global
momentum of the diagram ${\bf p}$. The times $\tau_1$ and $\tau_2$
are drawn from normalized probability distributions
$W_\Gamma(\tau_1)$ and $W_\downarrow(\tau_2)$ (arbitrary at this
point). Note that $W_\Gamma(\tau_1)$ and $W_\downarrow(\tau_2)$
can depend on ${\bf p}$ as a parameter. The acceptance ratio for
this update is
\begin{equation}
P_{\rm ins} = NC_{N+1} \, {\Gamma(\tau_1, {\bf p}) \,
G^{(0)}_\downarrow(\tau_2, {\bf p} )\over W_\Gamma(\tau_1)\,
W_\downarrow(\tau_2)}
 \; , \label{P_ins}
\end{equation}
where $C_N$ is an artificial weighing factor ascribed to
all worm diagrams of order $N$ (it can be used
for optimization purposes and depend on ${\bf p}$ too).
A natural choice for $W$-functions is
to make them proportional to BBL, i.e.
\begin{equation}
W_\Gamma(\tau) = {\Gamma(\tau, {\bf p}) \over \int
\Gamma(\tau', {\bf p})d\tau'}
 \,,\;W_\downarrow(\tau) = {G^{(0)}_\downarrow(\tau, {\bf p}) \over
\int  G^{(0)}_\downarrow (\tau', {\bf p}) d\tau'}
 \,  . \label{W}
\end{equation}
Then, to have an acceptance ratio of order unity and independent of
${\bf p}$ we choose
\begin{equation}
C_N = \frac{1}{N \Lambda}\,, ~~\Lambda = \int \Gamma(\tau_1, {\bf
p})d\tau_1 \int  G^{(0)}_\downarrow (\tau_2, {\bf p}) d\tau_2 \, .
\label{C_N}
\end{equation}
In the rest of the paper, we will refer to this choice of $W$ and
$C_N$ as ``optimized'', though we do not mean that it is the best
one possible for the entire scheme. For the ``optimized'' choice
\begin{equation}
P_{\rm ins} = N/(N+1) \; . \label{P_ins_opt}
\end{equation}

In the molecule sector, we essentially repeat all steps, up to
minor modifications. Now the propagator being selected is $\Gamma$
(once again, the update is rejected if the selected propagator is
covered.) Then, a pair of new propagators,
$G^{(0)}_\downarrow(\tau_1,{\bf p})$ and $\Gamma(\tau_2, {\bf
p})$, is inserted in front of the selected uncovered propagator.
The new propagator $\Gamma(\tau_2, {\bf p})$ inherits the outgoing
spin-up line previously connected to the selected $\Gamma$; the
latter gets instead the ${\cal M}$-end of the worm while the
${\cal I}$-end is attached to the new $\Gamma$. The acceptance
ratio is identical to (\ref{P_ins}) [for the optimized choice it
is $N/(N+1)$]. The polaron and molecule sectors are mutually
exclusive due to particle conservation, thus only one type of the
{\it Insert} update is applicable to a given diagram.

 {\bf Delete:} This update converts worm diagrams to
physical ones while reducing the diagram order. It applies only to
diagrams of order $N>1$ with worms being separated by one
uncovered BBL. Otherwise, the update is rejected. If worms are
separated by an uncovered BBL, then the left and right neighbor
BBLs are also uncovered, and an update opposite to {\it Insert} is
possible.  In {\it Delete} we simply remove two consecutive BBL
and worms from the diagram. Its  acceptance probability is the
inverse of Eq.~(\ref{P_ins}),
\begin{equation}
P_{\rm del} = {1\over (N-1)C_N}\,
{W_\Gamma(\tau_1)\, W_\downarrow(\tau_2) \over \Gamma(\tau_1, {\bf
p}) \, G^{(0)}_\downarrow(\tau_2, {\bf p} ) }
 \; . \label{P_del}
\end{equation}
With Eqs.~(\ref{W}), (\ref{C_N}), we have
\begin{equation}
P_{\rm del} = N/(N-1) \; . ~~~~~~~~~~~~~~{\rm (optimized)}
\label{P_del_opt}
\end{equation}

 {\bf Close:} The update applies only to diagrams
with the worms. The proposal is to connect ${\cal I}$ and ${\cal M}$ with a line
$\tilde{G}(\tau, {\bf q} )$ and eliminate worms from the diagram.
The momentum variable ${\bf q}$ is drawn from the probability distribution
$W_\uparrow ({\bf q})$, while $\tau$ is the time interval between
${\cal I}$ and ${\cal M}$
to be covered by the new propagator.
There are two ways of connecting ${\cal I}$ and ${\cal M}$, forwards and
backwards. The ambiguity is automatically resolved by the absolute
value of the momentum variable ${\bf q}$: if $q\geq p_F$
($q< p_F$), the propagator is supposed to go forwards (backwards).
In practice, we first select the direction (with equal probabilities),
and then generate the momentum variable $q$ accordingly:
either $q\geq p_F$ or $q<p_F$.

The acceptance ratio for this update is
\begin{equation}
P_{\rm cl} = {2 \over (2\pi )^3 NC_N} \, {\tilde{G} (\tau, {\bf
q})\over W_\uparrow ({\bf q})}\, \prod_\nu {D_\nu (\tau_\nu , {\bf
p}_\nu') \over D_\nu (\tau_\nu , {\bf p}_\nu )}
 \; , \label{P_cl}
\end{equation}
where the subscript $\nu$ runs through all BBLs to be covered by
the new propagator (clearly, $\tau=\sum_\nu \tau_\nu$).
Primes indicate new values of the corresponding momenta:
\begin{equation}
{\bf p}_\nu'  = {\bf p}_\nu - {\bf q} \; .  \label{p_prime_1}
\end{equation}
As usual, the distribution function $W_\uparrow ({\bf q})$ can depend
on $\tau$ and the direction of the propagator.
The natural choice for this function would be
\begin{equation}
W_\uparrow ({\bf q}) ={\tilde{G} (\tau, {\bf q})
\over \Omega(\tau)}  \, , \label{W_u_opt}
\end{equation}
\begin{equation}
\Omega(\tau) = \left\{
\begin{array}{l}
\int_{q\geq
p_F} \tilde{G}_\uparrow (\tau, {\bf q}) \, d{\bf q} \,~~{\rm (forward)} \; , \\
\int_{q<p_F} \tilde{G}_\uparrow (\tau, {\bf q}) \, d{\bf q}
\,~~{\rm (backward)} \; ,\end{array} \right. \label{Omega}
\end{equation}
leading to the optimized acceptance ratio
\begin{equation}
P_{\rm cl} = \frac{2 \Lambda \Omega(\tau)}{(2\pi )^3 }   \prod_\nu {D_\nu
(\tau_\nu , {\bf p}_\nu') \over D_\nu (\tau_\nu , {\bf p}_\nu )}
 \; .  \label{P_cl_opt}
\end{equation}

In this Section we deal with diagrams based on bare propagators.
To avoid double-counting, we have to exclude all cases which can
be reduced to ladders already summed in $\Gamma$'s. When ${\cal
M}$ and ${\cal I}$ are on the nearest-neighbor BBL the proposal to
connect them with the spin-up particle propagator is rejected. The
last rule to be monitored it to restrict all physical diagrams to
be either in the polaron or molecule sectors, i.e. sectors
different from 0 and 1 are not allowed.

 {\bf Open:} The update applies only to physical diagrams
and proposes to create a worm by selecting at random and removing
one of the spin-up propagators. The
acceptance ratio is given by the inverse of  (\ref{P_cl}), (\ref{P_cl_opt})
\begin{equation}
P_{\rm op} =  {(2\pi)^3NC_N\over 2} \, {W_\uparrow ({\bf q}) \over
\tilde{G}_\uparrow (\tau, {\bf q}) }\, \prod_\nu {D_\nu (\tau_\nu
, {\bf p}_\nu') \over D_\nu (\tau_\nu , {\bf p}_\nu )}
 \; , \label{P_op}
\end{equation}
where the subscript $\nu$ runs through all BBLs covered by the
propagator, $\tau=\sum_\nu \tau_\nu$,  and ${\bf q}$ is
the momentum of the selected spin-up propagator.
Primes indicate new values of the BBL momenta:
\begin{equation}
{\bf p}_\nu'  = {\bf p}_\nu + {\bf q} \; .  \label{p_prime_2}
\end{equation}
In the optimized version, we have
\begin{equation}
P_{\rm op} =  {(2\pi)^3\over 2 \Lambda \Omega(\tau) } \, \prod_\nu
{D_\nu (\tau_\nu , {\bf p}_\nu') \over D_\nu (\tau_\nu , {\bf
p}_\nu )}
 \; .   \label{P_op_opt}
\end{equation}

{\bf Reconnect:} The purpose of this update is to change the
topology of diagrams with the worm. The proposal is to select at
random one of the $\tilde{G}$-propagators and swap its outgoing
end with ${\cal M}$; the momentum of the propagator remains the
same, only its time variable changes from $\tau_0$ to $\tau_0'$.
The acceptance ratio is given by:
\begin{equation}
P_{\rm rec}\; =\;  {\tilde{G}(\tau_0', {\bf q})\over
\tilde{G}(\tau_0, {\bf q}) }\, \prod_\nu {D_\nu (
\tau_\nu , {\bf p}_\nu ') \over D_\nu ( \tau_\nu, {\bf p}_\nu)}
 \; . \label{P_rec}
\end{equation}
The subscript $\nu$ runs through all BBLs that will change their
momenta (and cover numbers $N_c$'s) as a result of the update.
Topologically, there are two different situations (the two a complementary
to each other in terms of the update): (i) ${\cal M}$
is covered by the propagator in question, and (ii) ${\cal M}$ is
not covered by the propagator.
Correspondingly, the new values of the diagram variables are
calculated as
\begin{equation}
(\tau_0', {\bf p}_\nu') = \left\{
\begin{array}{l}
(\tau_0-\tau,\, {\bf p}_\nu+{\bf q} )~~~~{\rm (i)} \; , \\
(\tau_0+\tau,\, {\bf p}_\nu-{\bf q} )~~~~{\rm (ii)} \;
,\end{array} \right. \label{new_time}
\end{equation}
with $\tau=\sum_\nu \tau_\nu$.

The above set of updates is ergodic. It can be supplemented by
additional updates that may improve the algorithm performance by
more efficient sampling of the diagram variables and topologies.
Over-complete sets of updates are also useful for meaningful tests
of the detailed balance. The possibilities are endless, and here
we simply mention two updates we have been using.

 {\bf Time shift:} We propose new time variables, $\tau_\nu \to
\tau_\nu'$, for all uncovered BBLs (labeled here with the subscript
$\nu$). The acceptance probability is
\begin{equation}
P_{\rm sh} = \prod_\nu {W_\nu (\tau_\nu, {\bf p})
\, D_\nu (\tau_\nu ', {\bf p} ) \over
W_\nu (\tau_\nu' , {\bf p}) \, D_\nu (\tau_\nu  , {\bf p}) }
\; . \label{P_br}
\end{equation}
All uncovered propagators have the same momentum ${\bf p}$. With
the optimized choice for $W_\nu (\tau_\nu , {\bf p}) \propto D_\nu
(\tau_\nu  ,{\bf p})$ the acceptance ratio becomes unity.

 {\bf Redirect:} Here we propose to select at random
one of the $\tilde{G}$-propagators and change its direction to the
opposite. Simultaneously, we change the momentum of the selected
propagator (resulting in new momenta for all BBL it covers or will
cover as a result of the update). Let the selected propagator be
$\tilde{G}(\tau, {\bf q})$, and the new one be $\tilde{G}(\tau',
{\bf q}')$ with $\tau=\sum_\nu \tau_\nu$,  and $\tau'=\sum_\lambda
\tau_\lambda $, where $\nu$ runs through all BBLs covered by the
propagator $\tilde{G}_\uparrow (\tau, {\bf q})$ and $\lambda $
runs through all BBLs to be covered by the propagator
$\tilde{G}_\uparrow (\tau', {\bf q}')$. We assume that ${\bf q}'$
is drawn from the distribution $W_\uparrow$ introduced above. In
this case, the acceptance ratio is given by
\begin{eqnarray}
P_{\rm rdr} = &&{ W_\uparrow (\tau, {\bf q})\,
\tilde{G}(\tau', {\bf q}')\over W_\uparrow (\tau', {\bf
q}')\, \tilde{G}(\tau, {\bf q}) }  \times  \nonumber \\
&& \left[ \prod_\nu {D_\nu (\tau_\nu, {\bf p}_\nu') \over D_\nu
(\tau_\nu, {\bf p}_\nu)} \right] \left[  \prod_\lambda {D_\lambda
(\tau_\lambda, {\bf p}_\lambda') \over D_\lambda (\tau_\lambda,
{\bf p}_\lambda)}\right]
 \; , \label{P_rdr}
\end{eqnarray}
\begin{equation}
{\bf p}_\nu' = {\bf p}_\nu+ {\bf q}\, , ~~~~~~~~{\bf
p}_\lambda' = {\bf p}_\lambda - {\bf q}' \; . \label{003}
\end{equation}
For the optimized choice of $W_\uparrow$, see Eq.~(\ref{W_u_opt}),
\begin{equation}
{ W_\uparrow (\tau, {\bf q}) \, \tilde{G}(\tau', {\bf
q}')\over W_\uparrow (\tau', {\bf q}')\, \tilde{G}(\tau,
{\bf q}) } \; \to \; {\Omega(\tau') \over \Omega(\tau)} \; .
\label{P_rdr_opt}
\end{equation}

{\bf Diagram sign:} The sign of a diagram with worms is somewhat
arbitrary since it is not physical. The ambiguity is resolved by
assuming that ${\cal M}$ is always connected to ${\cal I}$ in the
backward direction by an auxiliary unity propagator. Then, to
comply with the diagrammatic rules, we change the configuration
sign each time any of the following updates are accepted: (i)
Reconnect, (ii) Open/Close updates dealing with spin-up
propagators in the forward direction, (iii) Insert/Delete in the
molecule sector, and (iv) Redirect. For Open/Close updates dealing
with spin-up propagators in the backward direction the sign
remains the same, because here the sign coming from changing the
number of closed spin-up loops is compensated by the sign in
Eq.~(\ref{tilde_G}); the same is also true for Insert/Delete
updates in the polaron sector (due to our choice of the auxiliary
propagator sign). For precisely the same sign compensation reason,
the Redirect update {\it does} change the sign despite the fact
that it preserves the number of loops.

\begin{figure}[tbp]
\centerline{\includegraphics [bb=80 360 430 450,
width=\columnwidth]{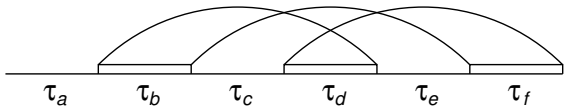}} \caption{A $G\Sigma$-diagram
contributing to the polaron self-energy. It factorizes into a
product of $G^{(0)}_{\downarrow}(\tau_a)$  and
$\Sigma(\tau=\tau_b+\tau_c+\tau_d+\tau_e+\tau_f)$.}
\label{G_sigma}
\end{figure}
\begin{figure}[tbp]
\centerline{\includegraphics [bb=80 360 430 450,
width=\columnwidth] {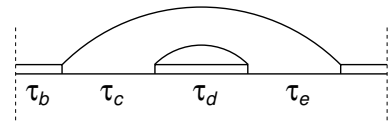} } \caption{A $\Gamma
\tilde{K}$-diagram contributing to the molecule self-energy. It
factorizes into a product of $\Gamma(\tau_b)$ and
$\tilde{K}(\tau=\tau_c+\tau_d+\tau_e)$. The vertical dashed lines
cut (for the sake of better visual perception) the same
$\Gamma(\tau_b)$-line.} \label{Gamma_K}
\end{figure}

\subsection {Estimators}
Only physical diagrams with one uncovered $G^{(0)}_{\downarrow}$
propagator contribute to the polaron self-energy. We will call
them $G\Sigma$-diagrams. An example is shown in
Fig.~\ref{G_sigma}. [Depending on restrictions imposed on the
configuration space (easy to implement in any scheme) physical
diagrams with more than one uncovered  $G^{(0)}_{\downarrow}$
propagator are either filtered out at the time when the
contribution to the self-energy histogram is made, or, are not
produced in the simulation at all.] They factorize into a product
of $G^{(0)}_{\downarrow}$ and some diagram contributing to the
self-energy $\Sigma$. The utility of cyclic representation is that
the uncovered propagator can be anywhere on the backbone. In view
of factorization, it is easy to write the MC estimator for the
integral (to simplify notations we omit irrelevant to the
discussion momentum ${\bf p}$)
\begin{equation}
I = \int_0^{\infty} f(\tau)\, \Sigma(\tau)\, d\tau \; ,
\label{I}
\end{equation}
where $f(\tau)$ is some function [see, e.g., Eqs.~(\ref{EE}),
(\ref{AA})]. Indeed, consider the estimator
\begin{equation}
{\cal E}_I\,  =\, \delta_{G\Sigma}\, f(\tau) \; . \label{EI}
\end{equation}
which counts all instances of $G\Sigma$-diagrams with an
additional weight $f(\tau)$.
Here $ \delta_{G\Sigma}$  is unity for each $G\Sigma$-diagram and zero
otherwise, and $\tau$ is the total duration in time of the
$\Sigma$-part of the $G\Sigma$-diagram. The Monte
Carlo average of this estimator is
\begin{equation}
\langle\, {\cal E}_I \, \rangle_{\rm MC}\, \propto\,
\int_0^{\infty} G^{(0)}_{\downarrow}(\tau')\, d\tau'
\int_0^{\infty} f(\tau) \Sigma(\tau)\,  d\tau  \, . \label{EI_MC}
\end{equation}
Similarly, within the same scheme, we collect statistics
of all ``normalization'' diagrams, see Fig.~\ref{zeroth}, using
an estimator projecting to the first-order diagram with the worm,
$\delta_{\rm norm}$. Then,
\begin{equation}
\langle\, \delta_{\rm norm} \, \rangle_{\rm MC}\, \propto\, C_1
\int_0^{\infty} G^{(0)}_{\downarrow}(\tau')\, d\tau' \,
\int_0^{\infty} \Gamma(\tau )\, d\tau  \; . \label{delta0_MC}
\end{equation}
The proportionality coefficient cancels in the ratio of the two
averages, leading to
\begin{equation}
I\, =\,  C_1{ \langle\, {\cal E}_I \, \rangle_{\rm MC}\over
\langle\, \delta_{\rm norm} \, \rangle_{\rm MC} }\,
\int_0^{\infty} \Gamma(\tau)\, d\tau \; . \label{est_I}
\end{equation}
In particular, for the optimal choice of $C_N$, we have
\begin{equation}
I\, =\,  { \langle\, {\cal E}_I \, \rangle_{\rm MC}\over \langle\,
\delta_{\rm norm} \, \rangle_{\rm MC} }\,\left[  \int_0^{\infty}
G^{(0)}_{\downarrow}(\tau)\, d\tau \right]^{-1} \; .
\label{est_I_opt}
\end{equation}

Imaginary-time integrals for the product of $\Sigma (\tau)$ and
exponentials, see Eqs.~(\ref{EE}) and (\ref{AA}), is all we need
to determine the polaron energy and residue. For the bold-line
generalization of the scheme described in the next Section we need
the entire dependence of self-energy on time and momentum. This is
achieved by differentiating partial contributions of
$G\Sigma$-diagrams. For example,
\begin{equation}
{\cal E}_{\Sigma,i}\,  =\, \delta_{G\Sigma}\; \delta_{\tau \in
{\rm bin}_i} \;, \label{ESt}
\end{equation}
is an estimator counting contributions with $\tau$ within the
$i$-th imaginary-time bin of width $\Delta \tau_i$ centered at the
point $\tau_i$. Due to linear relation between $I$ and
$\Sigma(\tau)$ we immediately realize that (for optimized choice)
\begin{equation}
\Sigma (\tau_i ) \,  = \,  { \langle\, {\cal E}_{\Sigma,i}  \,
\rangle_{\rm MC}\over \langle\, \delta_{\rm norm} \, \rangle_{\rm
MC} }\,\left[ \Delta \tau_i
 \int_0^{\infty}
G^{(0)}_{\downarrow}(\tau)\, d\tau \right]^{-1} \; . \label{est_St}
\end{equation}

In complete analogy with the polaron case, we consider $\Gamma
\tilde{K}$-diagrams that contain one, and only one, uncovered
$\Gamma$-propagator, see Fig.~\ref{Gamma_K}, and use them to
collect statistics for the molecule self-energy. Up to
straightforward replacements $G^{(0)}_{\downarrow}\leftrightarrow
\Gamma$, $\Sigma \leftrightarrow  \tilde{K}$ all relations of this
subsection hold true.

\section{Simulating $T$-matrix $\Gamma(\tau,p)$ by bold diagrammatic Monte Carlo }
\label{sec:gamma}

\begin{figure}[tbp]
\centerline{\includegraphics [bb=80 380 510 440,
width=\columnwidth]{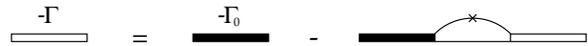}} \caption{Diagrammatic equation for the
$T$-matrix $\Gamma(\tau,p)$. The arc is the vacuum spin-up
propagator with the constraint $q<k_F$ on its momentum $q$. The
meaning of this equation is nothing but correcting the vacuum
result $\Gamma_0$ by subtracting contributions from the spin-up
fermions with momenta $q<k_F$.} \label{fig:Gamm}
\end{figure}

Despite relatively simple form of Eqs.~(\ref{Gamma}) and
(\ref{a}), tabulating the two-dimensional function $\Gamma (\tau,
p)$ with high accuracy using the inverse Laplace transform of
$\Gamma (\omega , p)$ turns out to be a time consuming job. In
this work we have used an alternative route based on the bold
diagrammatic Monte Carlo technology introduced recently in
Ref.~\cite{BMC}. The crucial observation is that the $T$-matrix
$\Gamma(\tau,p)$ can be diagrammatically related to its vacuum
counterpart $\Gamma_0(\tau,p)$, see Fig.~\ref{fig:Gamm}, with the
latter is known analytically:
\begin{equation}
\Gamma_0(\tau , p)\,  =\,  {4\pi \over m^{3/2}}
e^{(\epsilon_{F}+\mu-p^2/4m)\tau } g_{\mp}(\tau )\; ,
\label{Gammat2}
\end{equation}
were
\begin{equation}
g_{\mp}(\tau )\, =\,  -{1 \over \sqrt{\pi \tau}} \pm \sqrt{E} \;
e^{E \tau} {\rm erfc} (\pm \sqrt{E \tau} ) \; , \label{Gammat2a}
\end{equation}
for negative/positive values of the scattering length, $E=1/ma^2$,
and ${\rm erfc} (x)$ is the error-function. [The Fermi-energy and
the chemical potential in Eq.~(\ref{Gammat2}) come from the global
energy shift necessary for compliance with the Dyson equation shown in
Fig.~\ref{fig:Gamm}.]

With the explicit expression for the product of two vacuum
propagators the relation shown in Fig.~\ref{fig:Gamm} reads (the momentum
argument of $\Gamma (\tau, p)$ is suppressed for clarity)
\begin{eqnarray}
&-&\Gamma (\tau) \, = \, -\Gamma_0 (\tau)\, +\,  \int_0^{\tau} ds
\int_s^{\tau} ds' \:
\Gamma_0 (s) \, \Gamma (\tau - s')   \nonumber \\
&\times & \int_{q<k_F} {d{\bm q}\over (2\pi)^3} \left(-e^{-[({\bm
p -q})^2+{\bm q}^2](s'-s)/2m} \right) \; . \label{BG}
\end{eqnarray}

Equation (\ref{BG}) is one of the simplest examples of problems
solvable by bold diagrammatic Monte Carlo \cite{BMC}. We refer the
reader to Ref.~\cite{BMC}, where the algorithm of solving such
equations is described in detail. Here we just mention some
specific details. We find it helpful to start from a good trial
function for obtaining high-accuracy results in a short simulation
time. We achieve this goal using the following protocol. We start
the simulation by restricting imaginary time to be smaller than
$\tau_{\rm max}$ and select relatively short $\tau_{\rm max}$.
When the result is accurate enough, we extrapolate it to longer
times, increase $\tau_{\rm max}$, and restart the simulation with
the extrapolated function, $\Gamma_{\rm ext}$ serving as the trial
function, i.e. we substitute $\Gamma=\Gamma_{\rm ext}+\delta
\Gamma$ to Eq.~(\ref{BG}) and solve for $\delta \Gamma$. If
necessary, this procedure can be repeated several times.

\section{Simulation results and series convergence properties}
\label{sec:results}

Nearly all results in this paper were obtained by simulating
diagrams built on bare one- and two-particle propagators
$G^{(0)}_{\downarrow}$ and $\Gamma$. We observed that the
corresponding series are likely to be divergent. This, however,
does not mean that the entire idea of calculating contributions
from diagrams of higher and higher order and extrapolating results
to the infinite order is useless and ill-defined. On the contrary,
it was recognized long ago that appropriate re-summation
techniques allow one to determine reliably the function standing
behind the divergent series. Moreover, all re-summation techniques
(formally, there are infinitely many!), if applicable, have to
agree with each other on the final result. This important
observation vastly increases the utility of the Diag-MC technique
we are developing here. In the next Section we demonstrate that
making the series for $\Sigma$ self-consistent with the use of
Dyson equation---bold-line technique---is another way to improve
series convergence properties.

For the resonant Fermi-polaron considered here the Ces\`aro-Riesz
summation method solves the convergence problem. In general,
for any quantity of interest---in our case they are polaron or molecule
self-energy---one constructs partial sums
\begin{equation}
\Sigma (N_*)\, =\,  \sum_{N=1}^{N_*} D_N F_N^{(N_*)}\; ,
\label{partials}
\end{equation}
defined as sums of all terms up to order $N_*$ with the $N$-th
order terms being multiplied by the factor $F_N^{(N_*)}$. In the
limit of large $N_*$ and $N \ll N_* $ the multiplication factors
$F$ approach unity while for $N \to N_* $ they suppress
higher-order contributions in such a way that $\Sigma (N_*)$ has a
well-defined $N_* \to \infty$ limit. There are infinitely many
ways to construct multiplication factors satisfying these
conditions. This immediately leads to an important consistency
check: Final results have to be independent of the choice of $F$.
In the Ces\`aro-Riesz summation method we have
\begin{equation}
F_N^{(N_*)} = [(N_*-N+1)/N_*]^\delta \;, ~~~~~
\mbox{(Ces\`aro-Riesz)}\;. \label{factor1}
\end{equation}
Here $\delta > 0$ is an arbitrary parameter
($\delta = 1$ corresponds to the Ces\`aro method).
The freedom of choosing the value of Riesz's exponent $\delta$ can be
used to optimize the convergence properties of $\Sigma (N_*)$.
\begin{figure}[t]
\vspace*{-0cm}
\includegraphics[bb=0 15 270 210, width=\columnwidth]{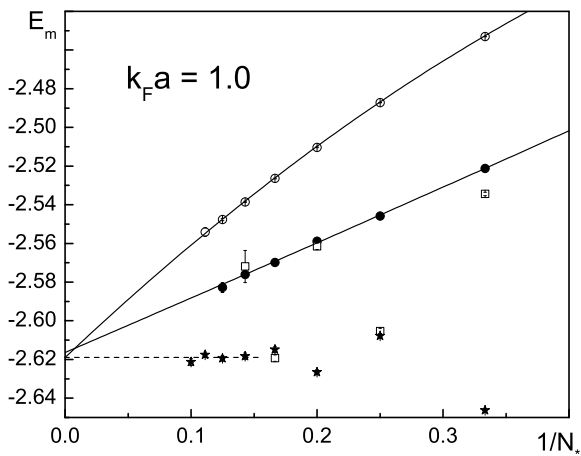}
 \caption{ The molecule  energy (at $k_F\, a=1$) as a function
of the maximum diagram order $N_*$ for different summation
techniques: Ces\`aro (open squares), Riesz $\delta=2$ (filled
circles, fitted with the parabola $y=-2.6164 + 0.28013x + 0.01638
x^2$), Riesz $\delta=4$ (open circles, fitted with the parabola
$y=-2.6190 + 0.61635 x - 0.3515 x^2$), and Eq.~(\ref{factor2})
(stars fitted with the horizontal dashed line). Reproduced from
Ref.~\cite{ourPRB}. } \label{molecule1}
\end{figure}
\begin{figure}[t]
\vspace*{-0cm}
\includegraphics[bb=0 0 580 800, angle=-90,  width=\columnwidth]{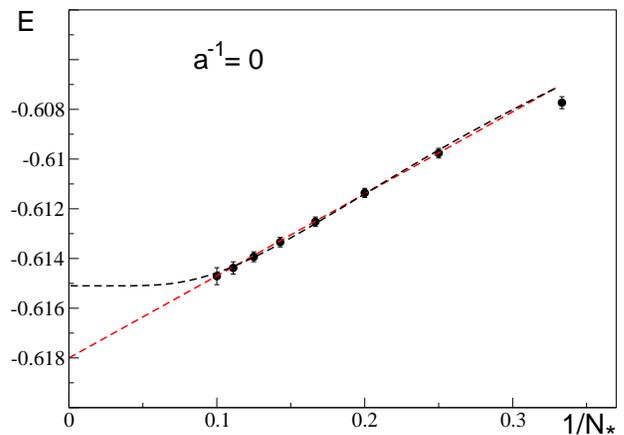}
 \caption{ (Color online)
The polaron energy (at the unitarity point $a^{-1}=0$) as a function
of the maximum diagram order $N_*$ using Eq.~(\ref{factor2}). The data are fitted
using linear $-0.618+0.033/N_*$ (red) and exponential $-0.6151+0.026e^{-0.39N_*}$
(black) functions to have an estimate of systematic errors
introduced by the extrapolation procedure. } \label{polaron_infty}
\end{figure}

We proceed as follows. For the series truncated at order $N_*$ we
first determine the polaron and molecule energies and then study
their dependence on $N_*$ as $N_* \to \infty$. In
Fig.~\ref{molecule1} we show results for the molecule energy at
$k_F\, a=1$. Without re-summation factors the data are oscillating
so strongly that any extrapolation to the infinite diagram order
would be impossible; we consider this as an indication that the
original series are divergent. Oscillations remain pronounced for
$\delta=1$, but are strongly suppressed for larger values of
$\delta$, so that for $\delta=4$ we were not able to resolve
odd-even oscillations any more. However, the smoothness of the
curve for large $\delta=4$ comes at the expense of increased
curvature, which renders the extrapolation to the $1/N_* \to 0$
limit more vulnerable to systematic errors. Empirically, we
constructed a factor $F_N^{(N_*)}$ which leads to a faster
convergence [see an example in Fig.~(\ref{molecule1})]:
\begin{equation}
F_N^{(N_*)} =\,
C^{(N_*)}\sum_{m=N}^{N_*}\exp{\left[-{(N_*+1)^2\over m(N_*-m+1)}
\right] } \; , \label{factor2}
\end{equation}
where $C^{(N_*)}$ is such that $F_1^{(N_*)}=1$. The most important
conclusion we draw from Fig.~\ref{molecule1} is that in our case
the series are subject to re-summation methods and the result of
extrapolation is method independent. We consider small variations
in the final answer due to different re-summation techniques and
extrapolation methods as
our systematic errors. An example is shown in
Fig.~\ref{polaron_infty}. In the next section we will present
evidence that the actual answer is closer to the upper bound of
$-0.615$. We see that in the absence of additional information one
has to allow for different ways of extrapolating the answer.

\begin{figure}[tb]
\vspace*{-0cm}
\includegraphics[bb=10 5 280 220, width=\columnwidth]{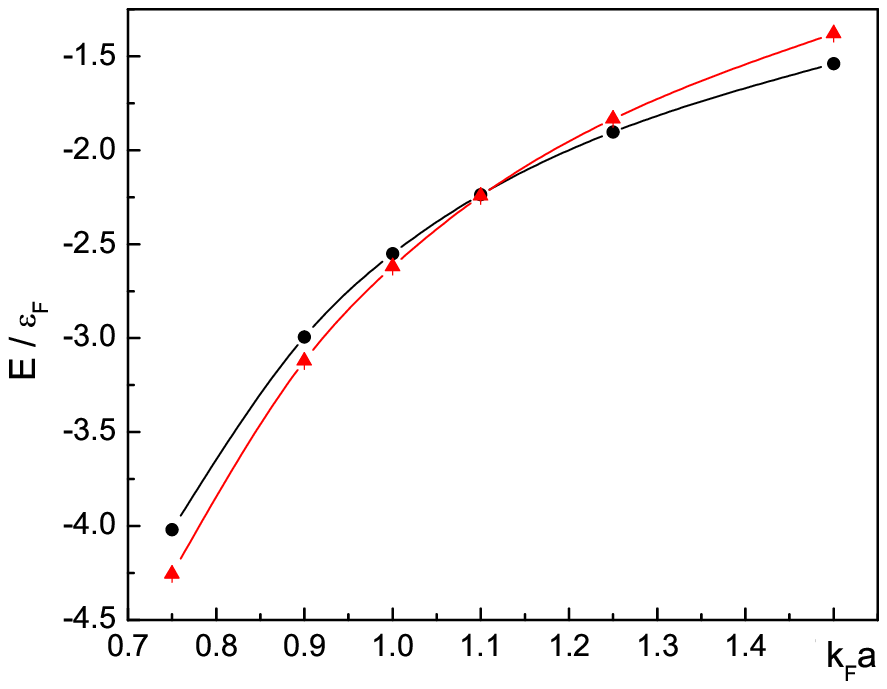}
\includegraphics[bb=10 15 280 210, width=\columnwidth]{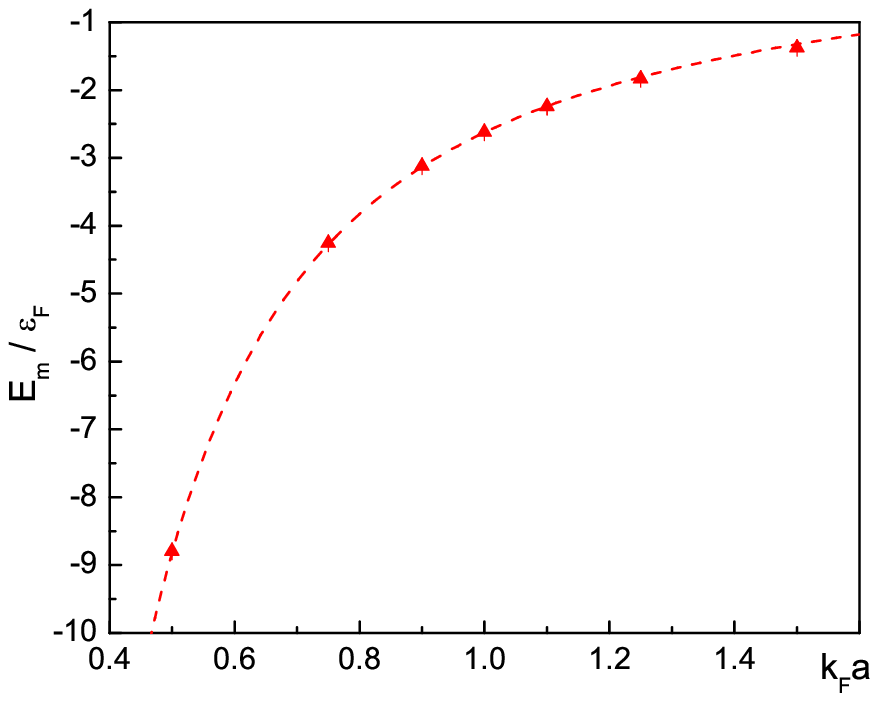}
 \caption{(Color online) Polaron (black circles) and molecule (red triangles) energies
 (in units of $\varepsilon_F$) as functions of $k_F a$.
 The dashed line on the lower panel corresponds to Eq.~(\ref{mol_asym}).
 Reproduced from Ref.~\cite{ourPRB}.}
 \label{skornyakov}
\end{figure}

Apart from consistency checks, one can test numerical results
against an analytic prediction for the strong coupling limit
$k_Fa\, \to 0$ corresponding to a compact molecule scattering off
majority spins. In this limit the molecule energy is given by the
expression
\begin{equation}
E_m =  -{1\over ma^2} - \varepsilon_F +  {2\pi
\tilde{a} \over (2/3)m } n_\uparrow ~~~~~~(k_Fa\, \to 0\, ) \, ,
\label{mol_asym}
\end{equation}
where the first term is the molecule binding energy in vacuum, the
second term reflects finite chemical potential of spin-up
fermions, and the last term comes from the interaction between the
composite molecule with the Fermi gas. The molecule-fermion
$s$-scattering length $\tilde{a} \approx 1.18 a$
\cite{Skornyakov,kagan} is obtained from the non-perturbative
solution of the three-body problem. Agreement with
Eq.~(\ref{mol_asym}) provides a robust test for the entire
numerical procedure of sampling asymptotic diagrammatic series.
Our data are in a perfect agreement with the $\tilde{a}\approx
1.18a$ result within the statistical uncertainty of the order of
$5\%$, see the lower panel in Fig.~\ref{skornyakov}.

Somewhat surprising outcome is that $E_m$ is described by
Eq.~(\ref{mol_asym}) very accurately all the way to the crossing point.
This fact can be used to approximate the energy density functional
of the superfluid polarized phase in the strongly imbalanced gas
for $k_Fa<1$ as that of the miscible dilute molecule gas coupled
to spin-up fermions \cite{Giorgini} (see also Refs.~\cite{Viverit,Pieri})
\begin{equation}
E={3\over 5} \epsilon_F n_{\uparrow} -\left[ {1\over ma^2} - {2\pi \tilde{a} \over (2/3)m } n_{\uparrow} \right]
n_{\downarrow} + {\pi a_{MM} \over m }  n_{\downarrow}^2 \;,
\label{functional}
\end{equation}
where $n_{\uparrow}$, $n_{\downarrow}$ are densities of unpaired spin-up fermions
and molecules, and $a_{MM}\approx 0.6a$ is the molecule-molecule
scattering length \cite{Petrov}. Within this approach it is
found that the system undergoes phase separation for $k_Fa > 0.56$
\cite{Giorgini}.

\begin{figure}[tbh]
\vspace*{-0cm}
\includegraphics[bb=15 15 270 220, width=\columnwidth]{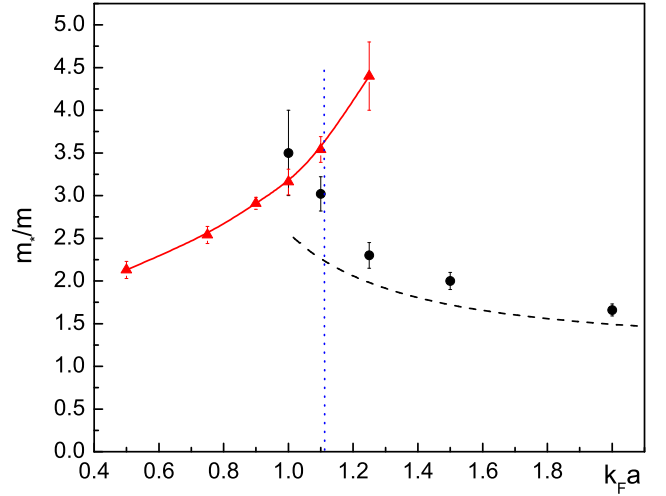}
 \caption{ (Color online) Polaron (black circles) and molecule
 (red triangles) effective
 mass as functions of $k_F a$.
 The vertical dotted line stands for $(k_F a)_c=1.11$.
 The dashed line  is the contribution from the first diagram
 \cite{Combescot}.
 Reproduced from Ref.~\cite{ourPRB}.} \label{masses}
\end{figure}

Phase separation precludes one from investigating the crossing
point between the polaron and molecule curves in trapped imbalanced
Fermi gases. It also rules out the multi-critical point on the
phase diagram predicted in Ref.~\cite{Sachdev}. This issues,
however, are not directly relevant to our study of properties of
one spin-down particle. In Fig.~\ref{skornyakov}, we present polaron
and molecule energies in the region of $k_Fa\sim 1$ where the
nature of the quasiparticle state changes. The crossing point is
found to be at $(k_Fa)_c = 1.11(2)$. Overall, both curves are in
excellent agreement with the variational Monte Carlo simulations
\cite{Lobo,Giorgini}. There is a certain degree
of accidental coincidence in the fact that the polaron self-energy
is nearly exhausted by the first-order diagram considered in
Ref.~\cite{Combescot}, see also Fig.~\ref{polaron_infty}. As we show
in the next section, both second-order and third-order diagrams
make considerable contributions to the answer but they happen to nearly
compensate each other, i.e. Green's function renormalization and
vertex corrections have similar amplitudes and opposite signs.

The intersection of the polaron and molecule curves
can be determined very accurately because both solutions are
describing well-defined quasiparticles at the crossing point.
This is because matrix elements connecting the two branches
involve at least four particles and their on-resonance
phase volume is zero at $(k_Fa)_c$. Indeed,
the energy, momentum, and particle number (for each spin direction)
conservation laws dictate that polaron decays into molecule,
two holes and one spin-up particle (molecule decays into
polaron two spin-up particles and one hole). For this process
the final-state phase volume gets negligibly small as compared
to the energy difference $|E_p-E_m|$ at and in the vicinity of
the crossing point.

The data for the effective mass is presented in Fig.~\ref{masses}.
At the crossing point the effective mass curve is discontinuous as
expected for the exact crossing between two solutions. One
important observation is that good agreement with
Eq.~(\ref{mol_asym}) for molecule energy in the entire region
$k_Fa<1$ does not guarantee yet that the compact-boson
approximation is accurate in the same region if properties other
than energy are addressed. In this sense, even good agreement with
Eq.~(\ref{mol_asym}) has to be taken with a grain of salt since
(\ref{mol_asym}) assumes that the molecule mass is $2m$ and
independent of $k_Fa$. The actual effective mass is significantly
enhanced in the vicinity of $(k_Fa)_c$.

\section{Numeric realization of the ``bold-line'' trick}
\label{sec:bold}

The success of the Diag-MC method for fermi-polarons is, to large
extent, due to small error bars we have for the sign-alternating
sums of high-order diagrams. In general, it is expected that the
computational complexity of getting small error bars for
sign-alternating sums in the limit of large $N$ is exponential (or
factorial) in $N$ since it usually scales with the configuration
space volume. Any tricks that reduce the configuration space while
keeping the scheme exact are worth investigating. In fact, we have
already used one of such tricks above when we introduced $\Gamma$
summing up all ladder diagrams. As a result, the diagram order was
defined by the number of $G_{\downarrow}$-lines, not bare
interaction potential vertexes, and ladder-reducible diagrams were
excluded from the configuration space.

In this Section we go one step further and apply another method,
well-known in analytic calculations (but virtually never carried
out analytically to high-order; for first-order diagrams it is
known as the self-consistent Born approximation). It is called the
``bold-line'' trick. The relation between $G_{\downarrow}$,
$G_{\downarrow}^{(0)}$ and $\Sigma$ accounts for the infinite sum
of diagrams forming geometrical series. Now, if one-particle lines
in self-energy diagrams are representing exact Green's function
(in this case they are drawn in bold) then many diagrams have to
be excluded to avoid double counting. Namely, any structure which
can be interpreted locally as part of the Dyson equation for
$G_{\downarrow}$ has to be crossed out. Though formally the
diagram order is still defined by the number of $G_{\uparrow}$, it
is in fact representing a whole class of diagrams (up to infinite
order) in the original, or bare, terms. Clearly, the MC scheme is
now self-consistently defined and potentially even finite number
of ``bold'' diagrams can capture non-perturbative effects.

Recently, we have demonstrated that the bold-line trick
is compatible with Diag-MC and the corresponding scheme has
been termed the bold diagrammatic Monte Carlo (BMC) \cite{BMC}.
There are two routes for implementing the bold-line technique. The
first one is to arrange two (running in parallel) coupled Monte
Carlo processes: one sampling the series for the self-energy in
terms of exact propagators, and the other one sampling
propagators from the Dyson equation. The latter process is
essentially the same as the process we use for pre-calculating
$\Gamma$, with an important new feature that the self-energy
used in the sampling is permanently updated. The
second route is specific for Dyson-type equations which allow
trivial algebraic solution in momentum representation.
In the present work, we use the second route.

The implementation of the bold-line trick requires that we introduce
an update which changes the global momentum ${\bf p}$ of the diagram.
This update applies only to the simplest normalization diagrams, see
Fig.~\ref{zeroth}. The integrated weight of normalization diagrams
is given by $\Lambda ({\bf p})$, see Eqs.~(\ref{C_N}) and (\ref{delta0_MC}).
In the update, we select the new value for the global
momentum from the probability distribution $\Lambda ({\bf p})$ and
propose new time variables for the spin-down and pair propagators
from the optimized probability distributions $W_\Gamma(\tau_1, {\bf p})$ and $W_\downarrow(\tau_2, {\bf p})$. Since new variables are seeded using
the exact diagram weight the acceptance ratio is unity. In practice,
the modulus of the global momentum variable is defined on the discrete
set of points.

We start the simulation with
$G_{\downarrow}=G_{\downarrow}^{(0)}$, and collect statistics to
the momentum-time histogram of $\Sigma$ from bold-line diagrams.
After a certain number of updates, we perform fast Fourier
transform of $\Sigma(\tau, p)$ to obtain $\Sigma(\omega, p)$,
calculate $G_{\downarrow}(\omega, p)$ using Dyson equation, which
is then transformed back to $G_{\downarrow}(\tau, p)$. The
simulation proceeds as before with the $G_{\downarrow}(\omega, p)$
function being recalculated at regular time intervals to reflect
additional statistics accumulated in $\Sigma$. Obviously, the
self-consistent feedback present in the bold-line scheme at the
beginning of the simulation violates the detailed balance equation
each time the function $G_{\downarrow}$ is updated. Only in the
long simulation time limit when both $\Sigma$ and $G$ do not
change any more is the detailed balance satisfied.

The other point which requires special care is the treatment of
ladder-reducible diagrams. In the bold-line implementation we have
to allow ladder diagrams back, but each spin-down line in the
ladder-reducible structure now has to be understood as a
difference $G_{\downarrow}-G_{\downarrow}^{(0)}$. Indeed, ladder
diagrams included in $\Gamma$ are built on bare propagators and
thus have to be corrected for the difference between the bare and
exact propagators.

Finally, we apply the bold-line approach in the molecule channel
as well.  In fact, the scheme was designed to be identical in the
one- and two-particle sectors. Now, in all self-energy diagrams we
have to substitute $G_{\downarrow}^{(0)}$ for $G_{\downarrow}$ and
$\Gamma$ for $Q$, with both $G$ and $Q$ being periodically
recalculated to reflect additional statistics statistics
accumulated to the $\Sigma$- and $\tilde{K}$-histograms.
Correspondingly, diagrams which can be locally interpreted as part
of the Dyson equation in the molecule sector have to be excluded.
[Ironically, this means that ladder diagrams are not allowed once
again with the exception for the first-order diagram in
$\tilde{K}$ which ensures that ladders are built on
$G_{\downarrow}$, see previous paragraph.] The updates and
acceptance ratios do not change in the bold-line representation,
but in the optimized version the probability distributions
$W_{\downarrow}$ and $W_{\Gamma}$ (and their normalization
integrals) are now proportional to $G_{\downarrow}$ and $Q$, i.e.
they have to be changed each time the new solutions of the Dyson
equations are generated.

As discussed in Ref.~\cite{BMC}, the Monte Carlo procedure of solving
self-consistent equations is more robust and has better convergence properties
then standard iterations, especially for sign-alternating series. There are
additional tools to improve the efficiency and convergence, some are
self-explanatory. It definitely helps to start with the initial function
$G_{\downarrow}$ as close as possible to the actual solution
(the final answer should not depend on small variations of the initial
choice). For example, the simulation for a given value of $k_Fa$ may start
with the final solution for the neighboring $k_Fa$ point.
The initial  statistics has to be discarded or ``erased'' according to some
protocol. If analytic expressions in special cases are available,
e.g. in the perturbation theory or strong coupling limits, they can be
used to match numeric data and restrict the parameter range
probed in the simulation.

\begin{figure}[tbh]
\vspace*{-0cm}
\includegraphics[
bb=0 0 580 800, angle=-90, width=\columnwidth]{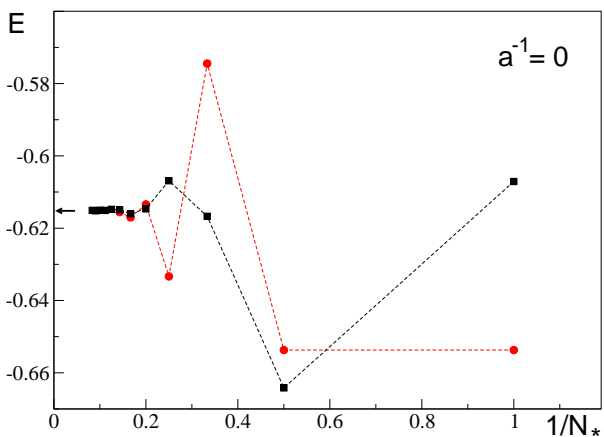}
 \caption{(Color online) Polaron energy as a function of the maximum diagram
 order $N_*$ at the unitary point $k_Fa=\infty$ within the bold-line
 approach. Black squares show results when the bold-line approach was
 implemented only for the polaron propagator. When both polaron
 and molecule propagators in diagrams are given by $G_{\downarrow}$ and
 $Q$ one obtains results shown by red circles.} \label{bold}
\end{figure}
In Fig.~\ref{bold} we present data for the polaron energy at the
unitary point calculated using the self-consistent scheme. As
before, the diagram order is defined by the number of spin-down
propagators. To see the difference between various approaches we
first calculated $E$ with the bold-line trick implemented only for
spin-down propagators (black squares) and then for both spin-down
and molecule propagators (red circles).  This plot makes it clear
that the perturbation theory result \cite{Combescot} is accurate
because corrections to spin-down propagators nearly cancel vertex
corrections. Figure~\ref{bold} is also telling the story which we
observe happening over and over again for other
strongly-correlated condensed matter problems: it does not really
make any sense to propose ``better'' approximations which account
only for some incomplete set of diagrams. For example, if in the
first-order diagram $\Gamma$ is replaced with $Q$ (self-consistent
Born approximation in the molecule sector), the answer is getting
much worse! Moreover, using exact expressions for $G_{\downarrow}$
and $Q$ in irreducible diagrams up to third-order (!) results in
an oscillation (the highest circle in Fig.~\ref{bold}) which forces one to
think that the final answer is probably even further up.
Fortunately, with the bold-line Diag-MC technique developed in
this article we can see how different approximations work and what
their actual value is, term by term.

\section{Conclusions}
\label{sec:colclusions}

The sign-problem in MC simulations is the problem of obtaining small error-bars
for system parameters which allow reliable extrapolation of results to the
thermodynamic limit. Most MC schemes are based on simulations of
finite-size systems (of linear size $L$) with the configuration
space volume growing exponentially/factorially with $L^3$ and inverse temperature
$ 1/T$ (for quantum models). Since error bars grow with the
configuration space volume they are completely out
of control {\it before} a meaningful extrapolation to the
thermodynamic limit can be done.

Computational complexity of the Diag-MC technique for
sign-alternating series is also exponential/factorial in the
diagram order and final results have to be extrapolated to the
$N_* \to \infty$ limit. In this sense the technique does not solve
the sign-problem, but offers a better route for handling it. One
important difference between the configuration space volume for
finite-size systems and connected Feynman diagrams is that the
latter deals with the thermodynamic limit directly. Moreover, the
same diagrams describe systems of different dimensions and
temperature. The list of advantages does not end here because one
can employ all known analytic tools to reduce the configuration
space, and thus make an exponential advance towards acceptable
solution of the sing-problem. By simulating the self-energy
instead of the Green's function (this was not done before) the
configuration space is reduced to that of $G$-irreducible graphs.
Using ladder diagrams we convert the standard perturbation theory
in the bare potential $V$ into the series expansion in terms of
$\Gamma$. Finally, the entire scheme is made self-consistent by
writing diagrams in terms of exact $G$ and $Q$. Since
self-consistency accounts for infinite sums of diagrams forming
geometrical series the configuration space of bold-line diagrams
is reduced further. All combined, the final formulation is compact
enough to perform the $N_* \to \infty$ extrapolation reliably
before error bars explode. Strictly speaking, having convergent
series is not a requirement because re-summation techniques are
well-defined mathematically and their work is guaranteed by
theorems based on properties of analytic functions.

At the moment we do not see any obvious limitations of the method
described here. On the contrary, we believe that it can be used
to study a generic interacting many-body system, Bose or Fermi.
Of course, the structure of diagrams and the optimal strategy
of applying analytic tools are Hamiltonian dependent and have to be
studied case-by-case. For example, in lattice models
there is no urgent need to deal with the ultra-violet limit
explicitly and one can proceed with the expansion in the bare
interaction potential $V$; the so-called
random phase approximation can be used to replace $V$ with the
screened interaction potential; the latter can be combined with
ladder diagrams, etc. The bold-line trick for Green's functions
can be implemented in any scheme.

To summarize, we have shown that polaron type problems can be studied
numerically with high accuracy using Diag-MC methods even when the
corresponding diagrammatic expansion is not sign-positive and divergent.
Previously such series were regarded as hopeless numerically, to such an
extent that nobody was actually trying! Using Diag-MC approach
we calculated energies and effective masses of resonant Fermi-polarons
in the BCS-BEC crossover region and determined that the point where the
groundstate switches from the single-particle (fermionic) sector to
two-particle (bosonic) sector is at $k_Fa = 1.11(2)$.
This point falls inside the phase separation region for the dilute
mixture of spin-down fermions in the Fermi gas of spin-up particles
\cite{Giorgini}.

We are grateful to C. Lobo, S. Giorgini, and R. Combescot for
valuable discussions and data exchange. The work was supported by
the National Science Foundation under Grants PHY-0426881 and
PHY-0653183.

\end{document}